\documentclass[twocolumn,showpacs,preprintnumbers,prl,amsmath,amssymb]{revtex4}

\usepackage{graphicx}
\usepackage{dcolumn}
\usepackage{bm}
\usepackage{braket}

\usepackage[T1]{fontenc}
\usepackage[latin1]{inputenc}

\begin{document}

\title{Universal enhancement of the optical readout fidelity of single electron spins}

\author{M.~Steiner}
\author{P.~Neumann}
\email{p.neumann@physik.uni-stuttgart.de}
\author{J.~Beck}
\email{j.beck@physik.uni-stuttgart.de}
\author{ F.~Jelezko}
\author{J.~Wrachtrup}

\affiliation{3. Physikalisches Institut, Universität Stuttgart, 70550 Stuttgart, Germany}

\date{\today}

\begin{abstract}
Precise readout of spin states is crucial for any approach towards physical realization of a spin-based quantum computer and for magnetometry with single spins. Here, we report a new method to strongly improve the optical readout fidelity of electron spin states associated with single nitrogen-vacancy (NV) centers in diamond. The signal-to-noise ratio is enhanced significantly by performing conditional flip-flop processes between the electron spin and the nuclear spin of the NV center´s nitrogen atom. The enhanced readout procedure is triggered by a short preparatory pulse sequence. As the nitrogen nuclear spin is intrinsically present in the system, this method is universally applicable to any nitrogen-vacancy center.
\end{abstract}

\maketitle


\indent The negatively charged nitrogen-vacancy (NV) center in diamond exhibits numerous outstanding properties which make it a promising candidate for novel applications in quantum and imaging science. Impressive experiments have demonstrated its potential as a solid-state qubit at room temperature \cite{jelezko-prl04,hanson-nature08,stoneham-physics09}, for nanoscale magnetometry \cite{gopi-nature08,maze-nature08,lukin-naturephys08} and for probing spin dynamics at nanoscale \cite{childress-science06,hanson-science08}. The coherence times of its electron spin are the longest reported for any solid-state system at room-temperature \cite{gopi-naturemat09,nori-prb09}. The use of nearby single nuclear spins as additional resource for quantum information \cite{wrachtrup-01} allows a variety of applications, from conditional quantum gates \cite{jelezko-prl04-2} and storage of quantum information to realization of small quantum registers and multi-partite entanglement \cite{neumann-science2008,lukin-science2007}. In all these experiments, the quantum information stored in the spin system is read out optically by recording spin-state dependent fluorescence rates of the NV center \cite{nizovtsev-physicaB2001,manson-prb06}. Even single nuclear spin qubits can be read out via the NV center by coherent mapping of the nuclear spin state onto the electron spin \cite{jelezko-prl04-2}.
Therefore, the overall readout fidelity of any application is limited by the signal-to-noise ratio of the optical readout process of the NV center.
Very recently significant fidelity enhancement was achieved by mapping the electron spin state onto $^{13}$C nuclear spins and repetitive readout \cite{lukin-science09}.\\
\indent In this Letter, we report a novel and universal method to enhance the signal-to-noise ratio of the optical readout process of NV center spins. To this end we exploit the spin dynamics of the intrinsic nitrogen nuclear spin that rely on a level-anticrossing (LAC) in the NV center´s excited state \cite{neumann-njp09,fuchs-prl08,jacques-prl09} to obtain a threefold enhancement in signal per readout step.
This speeds up the data acquisition process by a factor of 3 and corresponds to an increase of the signal-to-noise ratio by a factor of $\sqrt{3}$.
As the method exclusively utilizes interactions with the intrinsic nitrogen nucleus, its feasibility does not depend on the presence of further ancilla spins.\\
\begin{figure}[t]
	\includegraphics[width=8.5cm]{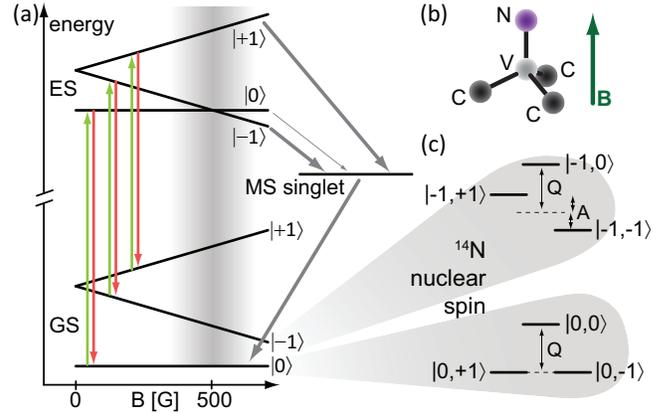}
	\caption{(a) Energy levels of the NV center as a function of the magnetic field amplitude B for \textbf{B} || NV symmetry axis. Optical transitions (vertical arrows) between ground (GS) and excited state (ES) are spin-conserving. Non-radiative ISC rates (dark grey arrows, thickness corresponds to transition rate) to the metastable singlet state (MS singlet) depend strongly on the spin state. (b) Atomic structure of the nitrogen-vacancy center in diamond and magnetic field orientation. (c) Detailed sublevel structure of the m$_{S}$=0 and m$_{S}$=-1 manifold, taking into account the $^{14}$N nuclear spin (I=1). Spin states are denoted by $\ket{m_{S},m_{I}}$. The hyperfine splitting is A = -2.166 $\pm$ 0.01 MHz. The nuclear quadrupole splitting Q has been measured to be 4.945$\pm$0.01 MHz by ENDOR spectroscopy \cite{clarke-textbook}. See supporting online material for details.}
	\label{fig1}
\end{figure}
\indent The NV center consists of a substitutional nitrogen atom and an adjacent vacancy (see Fig.\ \ref{fig1}(b)). Optical excitation of the transition between electronic ground and excited state gives rise to strong fluorescence which enables optical detection of individual NV centers by standard confocal microscopy techniques \cite{gruber-science97}. Ground and excited states are electron spin triplets (Fig.\ \ref{fig1}(a)). If the nitrogen atom is a $^{14}$N isotope (99,6\% abundance, I=1), each electron spin state is further split into three hyperfine substates (Fig.\ \ref{fig1}(c)). Optical cycles between ground and excited state are spin-conserving \cite{manson-prb06}. However, intersystem crossing (ISC) rates to an intermediate metastable singlet state are strongly spin-dependent. As the system cannot undergo optical cycles while being trapped in the singlet state ($\tau\approx$ 250 ns \cite{manson-prb06}), it remains dark during this time. Hence, the average fluorescence intensity depends on the spin state. ISC occurs mainly from the m$_{S}$=$\pm$1 ($\ket{\pm1}$) levels which therefore constitute "dark states", whereas the m$_{S}$=0 ($\ket{0}$) level constitutes a "bright state" with a higher average fluorescence intensity. ISC from the singlet state back to the ground state preferentially ends up in $\ket{0}$, leading to a strong polarization of the electron spin under optical excitation \cite{harrison-jlumin04}.\\
\indent The ground state spin triplet represents the logic qubit or the magnetic sensor. It is initialized to $\ket{0}$ by a nonresonant laserpulse (532 nm). The operating transition used in this Letter is the transition between $\ket{0}$ and $\ket{-1}$ (see Fig.\ \ref{fig2}(a)). Unitary qubit control or magnetic sensing is performed by common microwave pulse techniques under dark conditions. The output of the operation encoded in the spin state is read out optically by application of a readout laserpulse and detection of the fluorescence response. Fig.\ \ref{fig2}(a) shows the accumulated number of response photons per ns $n_{\ket{0}(\ket{-1})}(t)$ upon a readout laserpulse for the observed NV center initially being in spin state $\ket{0}(\ket{-1})$. If the initial spin state is $\ket{0}$, the pulse shows a high initial fluorescence level which decays towards a steady-state value with non-zero population in the singlet state due to a small probability for ISC from $\ket{0}$. For $\ket{-1}$, the initial fluorescence decays fast towards a low level due to a high ISC rate to the singlet state. As the singlet state always decays to spin state $\ket{0}$ in the ground state \cite{manson-prb06}, the low fluorescence level decays to the steady-state value within the lifetime of the singlet state of about 250 ns.\\
\indent The signal used to discriminate the spin states is the difference in the number of photons collected during the readout laserpulse (see grey area in Fig.\ \ref{fig2}(a)).
\begin{figure}[t]
	\centering
		\includegraphics[width=8.5cm]{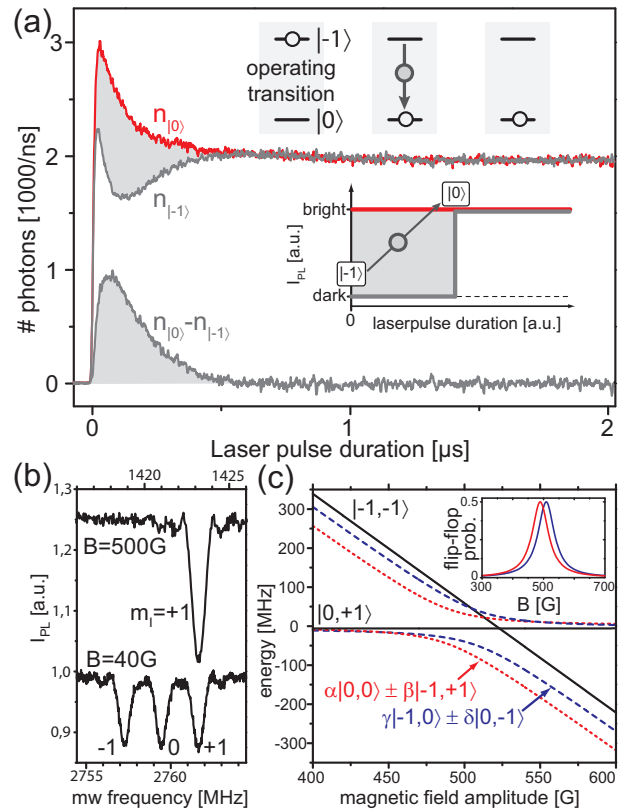}
	\caption{(color online) (a) Fluorescence responses for the system initially being in $\ket{0}$ (red trace n$_{\ket{0}}$) and $\ket{-1}$ (upper grey trace n$_{\ket{-1}}$) and their difference (lower grey trace). The grey area represents the signal that allows to discriminate different spin states. Inset: Schematic illustration of the signal formation. Starting in $\ket{-1}$, the system will pass through the dark singlet state (arrow with dot) and end up in the bright state $\ket{0}$ (steady state). A single passage through the singlet state yields the entire spin signal (grey area). (b) Spectra showing the $^{14}$N hyperfine structure of the $m_{S}$=-1 branch. At low field, the spin population is evenly distributed and three lines are visible. At B=500G, the nuclear spin is polarized into m$_{I}$=+1. (c) Excited state energy levels around B=500G. LAC occurs between $\ket{0,0}$ and $\ket{-1,+1}$ (red dotted lines) and $\ket{0,-1}$ and $\ket{-1,0}$ (blue dashed lines), enabling electron nuclear spin flip-flops. The inset shows the corresponding flip-flop probability per cycle through the excited state as a function of B.}
	\label{fig2}
\end{figure}
Note that for spin state $\ket{-1}$, the system passes once through the singlet state before being polarized. After polarization, all information about the initial spin state is destroyed and the system is in its steady state $\ket{0}$. Hence, the signal per readout pulse is limited by the optical polarization rate of the electron spin which is given by the lifetime of the singlet state ($\tau\approx$ 250 ns) and is on the order of 4 MHz. If fluorescence photons could be detected at a higher rate than 4 MHz, single passages through the singlet state could be observed as completely dark intervals in the fluorescence signal. However, as currently achievable photon countrates ($\approx$ 300 kHz at room temperature) are far below this threshold, readout has to be performed by repetitive accumulation of fluorescence signal with a concomitant increase in measurement time.\\
\indent We now demonstrate a method to decrease the measurement time (increase the number of signal photons per shot) for the spin state of single NV centers by a factor of 3 by exploiting the spin dynamics of the $^{14}$N nitrogen nuclear spin (I=1). We make use of a recently discovered nuclear spin polarization mechanism \cite{jacques-prl09} that is mediated by a level anti-crossing (LAC) in the excited state \cite{neumann-njp09}. Note that this mechanism has been demonstrated for a $^{15}$N nucleus (I=1/2), however, it works analogously for $^{14}$N. Spin states will be denoted by $\ket{m_{S},m_{I}}$ in the following.\\
\indent At a magnetic field of 500G (\textbf{B} || NV-axis), the m$_{S}$=0 and the m$_{S}$=-1 branch of the NV center electron spin are expected to cross in the excited state (see grey region in Fig.\ \ref{fig1}(a)). However, due to strong hyperfine coupling between electron and nitrogen nuclear spin in the excited state ($\approx$ 20 times stronger than in the ground state \cite{fuchs-prl08}), there is LAC between spin states $\ket{0,0}$ and $\ket{-1,+1}$ resp. $\ket{0,-1}$ and $\ket{-1,0}$, associated with strong spin mixing (see Fig.\ref{fig2}(c)). This allows energy-conserving flip-flop processes between electron and nuclear spin. These are not possible for spin states $\ket{0,+1}$ and $\ket{-1,-1}$ which therefore are not affected by mixing. Under optical illumination, the electron spin is steadily polarized into $\ket{0}$. Thus, the only stable spin state is $\ket{0,+1}$. As a result, optical illumination leads to strong polarization of the system into $\ket{0,+1}$ at B$\approx$500G (see spectra in Fig.\ \ref{fig2}(b)).\\
\indent If the system is now prepared in spin state $\ket{-1,-1}$ before application of the readout laserpulse, it has to pass three times through the singlet state instead of one in a cascade-like process before reaching the bright steady state $\ket{0,+1}$. As each passage through the singlet state yields the same amount of signal as obtained by conventional readout, the total signal is tripled.\\
\indent The detailed process occuring upon application of a readout laserpulse is illustrated in Fig.\ \ref{fig3}(a).
\begin{figure}[t]
	\centering
		\includegraphics[width=8.5cm]{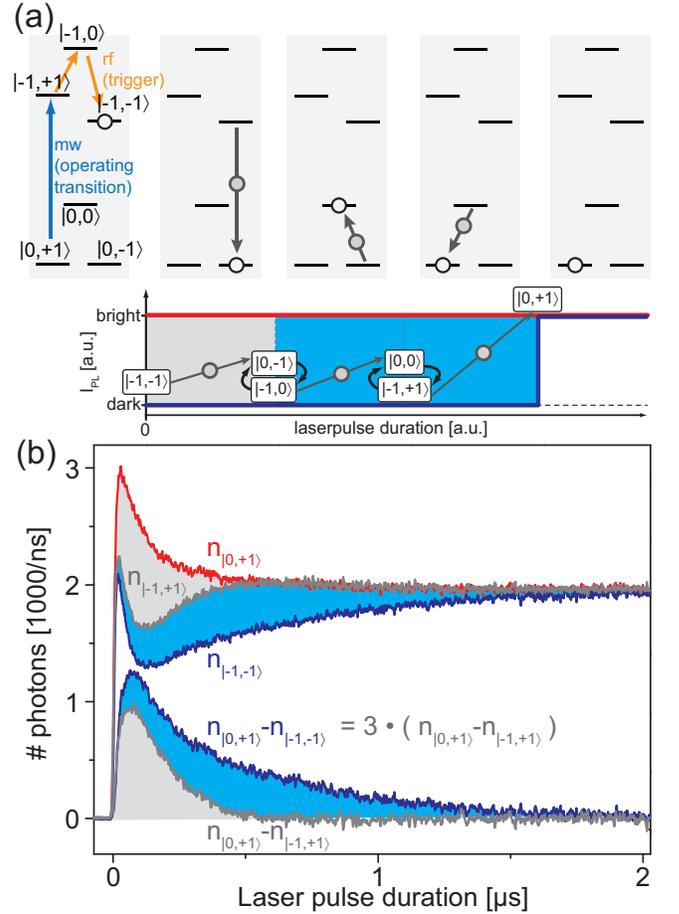}
	\caption{(color online) (a) Illustration of the enhanced readout process. The spin state $\ket{-1,-1}$ is populated by a selective microwave $\pi$-pulse (blue arrow) and two consecutive radiofrequency $\pi$-pulses (orange arrows). Due to electron nuclear spin flip-flops in the excited state, the system must pass through the dark singlet state (grey arrows with dots) three times before reaching the bright steady state $\ket{0,+1}$. The entire signal (grey and dark blue areas) is now three times the conventional signal (only grey area). (b) Fluorescence responses for the system initially being in $\ket{0,+1}$ (red trace n$_{\ket{0,+1}}$), $\ket{-1,+1}$ (conventional readout, grey trace n$_{\ket{-1,+1}}$) and $\ket{-1,-1}$ (enhanced readout, blue trace n$_{\ket{-1,-1}}$) and the corresponding differences (lower traces). By enhanced readout, the conventional readout signal (grey area) is increased by a factor of 3 (grey + dark blue area).}
	\label{fig3}
\end{figure}
Starting from spin state $\ket{-1,-1}$, the system once passes through the singlet state as the electron spin is polarized. This process yields the signal (grey area in Fig.\ \ref{fig3}(a)) obtained equivalently by conventional readout. The passage through the singlet state (grey arrows with dots in Fig.\ \ref{fig3}(a)) is assumed to conserve the nuclear spin state, hence the system ends up in $\ket{0,-1}$. The system is now repumped to the excited state by the same readout laserpulse. There it has a certain probability to perform the flip-flop process $\ket{0,-1}\leftrightarrow\ket{-1,0}$ due to the strong mixing between these two spin states (see Fig.\ref{fig2}(c)). Thus, it has the two possibilities of either performing an optical cycle under emission of a fluorescence photon or performing an electron-nuclear flip-flop process. As optical cycles are spin-conserving, a flip-flop process will finally occur and the system will be in $\ket{-1,0}$. From there, it will pass a second time through the singlet state, additionally giving rise to the same amount of signal as before (blue area in Fig.\ \ref{fig3}(a)). After the second relaxation via ISC, the system will be in state $\ket{0,0}$. It will now again be reexcited, where the spin states $\ket{0,0}$ and $\ket{-1,+1}$ are mixed. As before, the system will inevitably perform a spin flip-flop with subsequent passage through the singlet state, again yielding additional signal. After this third ISC-relaxation, the system will be in the bright steady state $\ket{0,+1}$, which yields a constant level of fluorescence intensity.\\
\begin{figure}[t]
	\centering
		\includegraphics[width=8.5cm]{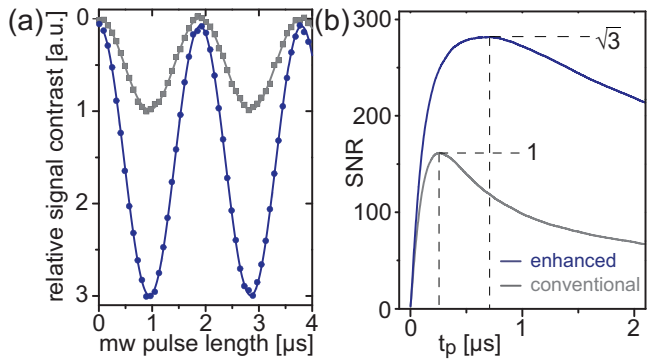}
	\caption{(color online) (a) Electron spin Rabi oscillations recorded by conventional (grey squares) and enhanced readout (dark blue dots). The relative contrast is improved by a factor of 3. (b) Signal-to-noise ratio ($SNR$) of the fluorescence responses from Fig.\ \ref{fig3}(b) for enhanced (dark blue) and conventional readout (grey) as a function of the readout pulselentgh t$_{p}$. The maximum $SNR$ for enhanced readout at B=500G is $\sqrt{3}$ times the maximum $SNR$ for conventional readout.}
	\label{fig4}
\end{figure}
\indent The fluorescence responses $n_{\ket{m_S,m_I}}(t)$ for conventional (grey trace) and enhanced readout (blue trace) are compared in Fig.\ \ref{fig3}(b). The lower traces show the difference in fluorescence between bright and dark state for both cases. The colored areas represent the amount of signal photons and show the expected threefold enhancement.\\
\indent The amount of signal photons saturates with increasing pulse duration while the noise arising from the poissonian distribution of collected photons (shot noise) grows approximately as the squareroot of the pulselength. Hence, there is an optimal readout pulselength which maximizes the signal-to-noise ratio ($SNR$). The signal acquired during the initial time interval $[0,t_{p}]$ of the fluorescence pulses is $N_{\ket{0,+1}}(t_{p})-N_{\ket{-1,m_I}}(t_{p})$ (grey area for m$_I$=+1 in Fig.\ \ref{fig3}(b) and grey + dark blue area for m$_I$=-1). $N_{\ket{m_{S},m_{I}}}(t_{p})$ is given by $\sum_{0}^{t_{p}}n_{\ket{m_{S},m_{I}}}(t)$. The shot noise is the squareroot of the total number of collected photons, $\sqrt{N_{\ket{0,+1}}(t_{p})+N_{\ket{-1,m_{I}}}(t_{p})}$. The $SNR$ is
\begin{equation}
SNR~(t_{p})=\frac{N_{\ket{0,+1}}(t_{p})-N_{\ket{-1,m_I}}(t_{p})}{\sqrt{N_{\ket{0,+1}}(t_{p})+N_{\ket{-1,m_I}}(t_{p})}}.
\end{equation}
It has a global maximum at an optimal readout pulselength $t_p$ (see Fig.\ \ref{fig4}(b)).\\
\indent For enhanced readout, both the signal and the time required for signal formation are increased by a factor of 3. Thus, the maximum $SNR$ is enhanced by $\sqrt{3}$ (see Fig.\ \ref{fig4}(b)) and shifted to a longer pulselength. Fig.\ \ref{fig4}(a) shows electron Rabi oscillations recorded with conventional and enhanced readout. The experimental results exhibit the behaviour predicted by our model.\\
\indent In practice, enhanced readout is implemented as follows: After execution of a desired pulse sequence on the operating transition $\ket{0,+1}\leftrightarrow\ket{-1,+1}$ (see arrow in Fig.\ \ref{fig3}(a)), two consecutive resonant radiofrequency $\pi$-pulses on the nuclear spin transitions $\ket{-1,+1}\leftrightarrow\ket{-1,0}$ and $\ket{-1,0}\leftrightarrow\ket{-1,-1}$ (see orange arrows in Fig.\ \ref{fig3}(a)) are applied right before application of the readout laserpulse. Nuclear spin transitions can be driven selectively up to Rabi frequencies on the order of MHz. Thus, the time required to trigger the enhanced readout is on the order of a few $\mu$s. At B=500G, the optimum pulselength for the enhanced fluorescence response is extended by about 500 ns. Hence, the enhanced readout method already pays if the length of the complete pulse sequence is on the order of a few $\mu$s, which is fulfilled in most cases.\\
\indent The most critical parameter of enhanced readout is the angle between magnetic field and the NV axis \cite{jacques-prl09}. The magnetic field can be varied over a wide range ($\pm$200G) without losing the polarization effect (see \cite{jacques-prl09}). However, as the flip-flop probability decreases with distance from the LAC (see inset in Fig.\ \ref{fig2}(c)), the signal formation process is slowed down which leads to decrease of the maximum $SNR$. At 50G from the LAC, the $SNR$ enhancement is about half the maximum value.\\  
\indent The method has been demonstrated for NV centers containing a $^{14}$N atom (I=1), however, it works analogously for $^{15}$N (I=1/2), where a $SNR$ enhancement of $\sqrt{2}$ is achievable by a single rf pulse. Note that this is the first demonstration of rf-control of a single nitrogen nuclear spin which proves its suitability as an additional qubit intrinsic to the NV center.\\
\indent Summarizing, we presented a fast universal method to enhance the signal-to-noise ratio of the optical readout process of the NV center in diamond.
The measurement time required to determine the spin state of the system is reduced by a factor of 3.
This allows for faster room-temperature access to quantum information stored in the individual solid-state spin system and speeds up data acquisition in spin-based magnetometry.
Further enhancement can be achieved by including those $^{13}$C nuclear spins that are polarized at magnetic fields corresponding to the excited state LAC \cite{jacques-prl09,childress-arx09}.

\newpage

\section{Additional material}
In this additional online material we explain the spin Hamiltonian and the important interactions in more detail. In addition the $^{14}$N nuclear spin spectra are presented. To record them the new signal enhancement method has been used.

\maketitle

\subsection{The NV Hamiltonian}
\begin{figure}[!t]
 \centerline{\includegraphics[width=8.0cm]{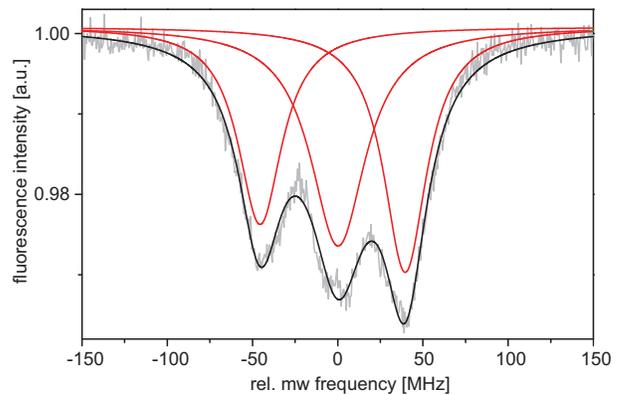}}
    \caption{Excited state electron spin resonance spectrum. The spectrum shows the $m_S=0 \leftrightarrow -1$ transition of the electronic excited state of the NV center. It is fitted with 3 Lorentzians (red lines) corresponding to the hyperfine splitted $^{14}N$ nuclear spin sublevels. The hyperfine interaction in the excited state is roughly 20 times larger than in the ground state. The width of the lines originates from the short excited state lifetime of about 10~ns.}
    \label{figS2}
    \end{figure}

For the NV center with a $^{14}$N isotope the spin Hamiltonian for the ground state is
\begin{equation}
H=D\hat{S}_z^2 + g_{e}\mu_{B} B \hat{S}_z + A \; \hat{\underline{S}} \; \hat{\underline{I}} + Q \hat{I}_z^2 + g_{n}\mu_{n} B \hat{I}_z
\label{Hamilto}
\end{equation}
where the first and second terms express electron spin energies with the zero field splitting $D=2870$ MHz of levels $m_S=0$ and $m_S=\pm1$ and the Zeeman energy with the electron g-factor $g_e$ and the Bohr-magneton $\mu_B$.
The third term is the hyperfine interaction between NV center's electron spin and the nuclear spin of its $^{14}$N atom.
As the electron spin the nuclear spin is a triplet $I=1$.
It leads to a splitting of $\approx2.2$~MHz.
For simplicity we assume an isotropic hyperfine interaction which is sufficient to explain the spectra in the paper.
Finally, the fourth and fifth term describe the energies of the nuclear spin with its quadrupole splitting $Q$ and its nuclear Zeeman energy where $g_n$ is the $^{14}N$ nuclear spin g-factor and $\mu_n$ the nuclear magnetic moment.
In this Hamiltonian the magnetic field is assumed parallel to the NV-axis (z-direction) as was the case in the performed experiments.\\
\indent The resulting electron spin energy levels as a function of the magnetic field are displayed in Fig.~1(a) of the main paper and their nuclear spin sublevels are depicted in Fig.~1(c).
The electron spin transition between states $\left|0\right.\rangle$ and $\left|-1\right.\rangle$ for two magnetic field strengths is shown in fig. 2(b) of the main paper.
In the lower spectrum the hyperfine splitting is visible.
It is absent in the upper spectrum because the nuclear spin is polarized in state $m_I=+1$.\\
\indent The excited state Hamiltonian basically looks the same.
Only the constants change.
The zero field splitting becomes $D_{es}=1420$~MHz \cite{Sneumann-njp09,Sfuchs-prl08}, and the hyperfine splitting is $A_{es}\approx40$~MHz (see Fig.~\ref{figS2}).
Again we assume an isotropic hyperfine interaction.
The fact that the polarization mechanism for the nuclear spin mentioned in the paper works well underlines that there must be substantial off-diagonal terms in the hyperfine interaction of the excited state.\\
\indent In Fig.\ 2(c) of the main paper, the calculated eigenvalues of the excited state Hamiltonian in the basis $\ket{-1,-1}$, $\ket{-1,0}$, $\ket{-1,+1}$, $\ket{0,-1}$,  $\ket{0,0}$ and $\ket{0,+1}$ are drawn as a function of the magnetic field amplitude.
The inset shows the probability for a spin flip-flop between electron and nuclear spin as a function of B. It has been calculated from the coefficients of the basis states participating in a flip-flop process (see \cite{Sjacques-prl09} for details).\\

\subsection{nuclear spin manipulation}
\begin{figure}[t]
 \centerline{\includegraphics[width=8.0cm]{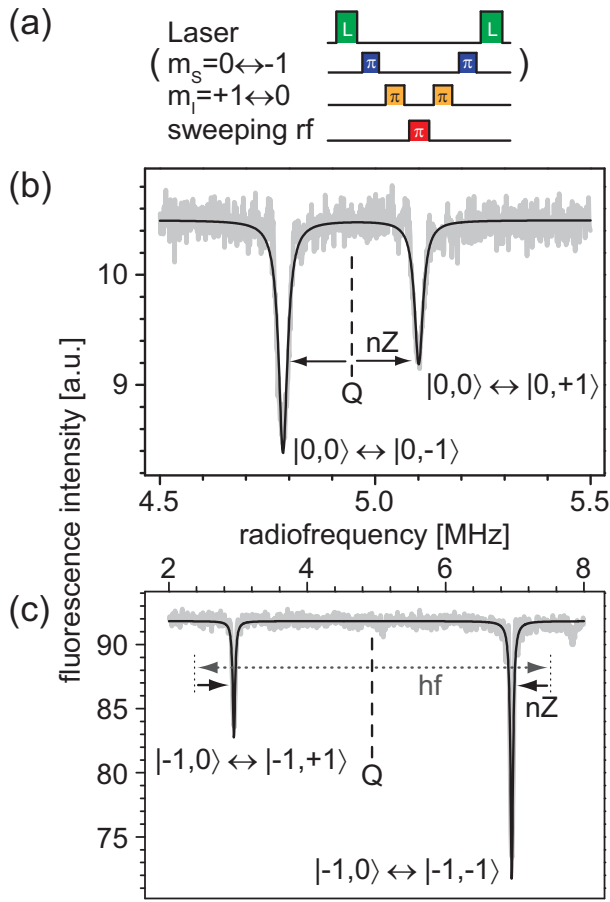}}
    \caption{\textbf{Single nuclear spin resonance} (a) Pulse sequence to obtain resonance spectrum of a single $^{14}N$ nuclear spin. The laser pulse initializes the system and also reads out the spin state. All microwave (mw) and radio frequency (rf) $\pi$ pulses with fixed frequency (blue and orange) do transformations towards and back from the starting state $m_I=0$. In between a $\pi$ pulse with sweeping frequency actually is responsible for the spectrum. (b) Nuclear spin spectrum for electron spin state $m_S=0$. The nuclear quadrupole splitting (Q) results in a frequency offset and the nuclear Zeeman energy (nZ) splits the two lines. (c) Spectrum for electron spin state $m_S=-1$. In addition to the quadrupole splitting and nuclear Zeeman energy the hyperfine interaction (hf) shifts the two lines. In all spectra, resonance lines are fitted with Lorentians.}
    \label{figS1}
    \end{figure}

For the enhanced readout technique, the nuclear spin has to be manipulated.
Therefore, first of all the nuclear transition frequencies in electron spin state $\left|0\right.\rangle$ and $\left|-1\right.\rangle$ have to be obtained.
This is done by a combined pulsed and cw technique which is sketched in Fig.~S\ref{figS1}(a).
The signal gained in this experiment already uses the enhancement technique.\\
\indent At first a magnetic field of 500 Gauss is applied parallel to the NV axis ($z$ direction) as in the paper.
A laser pulse initializes the system into $\left|m_S,m_I\right.\rangle=\left|0,+1\right.\rangle$.
A subsequent microwave (mw) $\pi$ pulse (\textit{not necessary for operation in $\left|\mathit{0}\right.\rangle$}) converts it into $\left|-1(\mathit{0}),+1\right.\rangle$.
In a next step a radiofrequency (rf) $\pi$ pulse resonant on transition $\left|-1(\mathit{0}),+1\right.\rangle \leftrightarrow \left|-1(\mathit{0}),0\right.\rangle$ yields state $\left|-1(\mathit{0}),0\right.\rangle$.
Out of this starting state a rf $\pi$ pulse with sweeping frequency is applied.
Finally population in $\left|-1(\mathit{0}),0\right.\rangle$ is converted back to $\left|0,+1\right.\rangle$ which is the only bright state.
The population here is read out by a final laser pulse which is in fact the first laser pulse of the next run.
This sequence is continuously repeated while the frequency of the sweeping pulse is changed and the corresponding fluorescence is recorded.
If the swept frequency hits the resonance for transitions $\left|-1(\mathit{0}),0\right.\rangle \leftrightarrow \left|-1(\mathit{0}),+1\right.\rangle$ or $\left|-1(\mathit{0}),0\right.\rangle \leftrightarrow \left|-1(\mathit{0}),-1\right.\rangle$ no population ends up in bright state $\left|0,+1\right.\rangle$.
It is then either in $\left|0,0\right.\rangle$ or in $\left|0,-1\right.\rangle$.
In this case the final laser pulse eventually forces the nuclear spin to flip and pass through the metastable state which gives the signal as explained in the paper.\\
\indent Note, that the electron spin state is already $m_S=0$ before the readout laser pulse so that it does not have to flip and does not contribute to the signal. This means that the nuclear spin state is read out without mapping it onto the electron spin due to the nuclear spin state selective flip-flop processes in the excited state. 
Be also aware, that the signal for spin state $\left|0,-1\right.\rangle$ is twice as high as the one for $\left|0,0\right.\rangle$ because the system has to pass the metastable state twice instead of once.\\
\indent The resulting nuclear spin spectra are shown in Fig.~S\ref{figS1}(b,c).
From there a quadrupole splitting of $Q=4.945\pm0.01$~MHz can be deduced and the hyperfine splitting is $A=-2.166\pm0.01$~MHz.
These values have been obtained for a few centers in different samples and have agreed within the given bounds.
Nevertheless, they show small deviation from earlier obtained values \cite{Smanson-joptb99}.\\ \\
The presented method is universal in the sense that the nitrogen nuclear spin is present for every NV center. For NV centers that have certain $^{13}$C nuclear spins in their vicinity, the signal-to-noise ratio can even be further enhanced. For $n$ participating nuclei with spin I$_n$, the enhancement is given by $SNR_{enh}/SNR_{conv}=\sqrt{1+{\sum_{n}} 2I_{n}}$. Note that this is true only for nuclear spins which have the same quantization axis as the NV center electron spin and which participate in energy conserving flip-flop processes mediated by the excited state LAC leading to nuclear spin polarization \cite{Schildress-arx09}.


\begin{thebibliography}{16}
\bibitem{quadrupole}
All experiments have been carried out in type IIa diamond samples.
\end{thebibliography}
\end{document}